\documentclass[sigconf, authorversion, nonacm]{acmart}
\usepackage{graphicx}
\usepackage{subcaption} 

\usepackage{booktabs}
\usepackage{multirow}

\AtBeginDocument{%
  }

\setcopyright{acmlicensed}
\copyrightyear{2018}
\acmYear{2018}
\acmDOI{XXXXXXX.XXXXXXX}
\acmConference[Conference acronym 'XX]{Make sure to enter the correct
  conference title from your rights confirmation email}{June 03--05,
  2018}{Woodstock, NY}
\acmISBN{978-1-4503-XXXX-X/2018/06}

\newcommand{\QUOTE}[1]{\textsf{\textit{\textcolor{black}{``#1''}}}}

\begin{document}

\title{Moodie: An Early-Stage Design Exploration for Supporting Fear of Missing Out with LLM-based Chatbots}


\author{Hsin-Yu Tsai}
\affiliation{%
  \institution{National Chung Cheng University}
  \country{Taiwan}}
\email{hsinyu0519@gmail.com}

\author{Jingxian Liao}
\affiliation{%
  \institution{UC Davis, Davis}
  \country{United States}}
\email{jxliao2021@gmail.com}

\author{Fu-Yin Cherng}
\affiliation{%
  \institution{National Cheng Kung University}
  \country{Taiwan}}
\email{fuyincherng@gs.ncku.edu.tw}

\author{Tzu-Hsiang Huang}
\affiliation{%
  \institution{National Chung Cheng University}
  \country{Taiwan}}
\email{tommy900916@gmail.com}

\renewcommand{\shortauthors}{Tsai et al.}

\begin{abstract}
The excessive use of social media has led to the challenge known as Fear of Missing Out (FoMO).
Existing studies fail to provide accessible, interactive tools that focus on the emotional and cognitive aspects of FoMO.
This work presents Moodie, a chatbot designed using Large Language Models to support emotion regulation and reduce FoMO.
We conducted a formative study to understand the needs of individuals with FoMO and developed Moodie.
Then, we conducted a preliminary evaluative study (N=21) to observe how participants interact with Moodie and a baseline chatbot (GPT-4o) over one week.
The results show that while both Moodie and a baseline chatbot reduced FoMO to a similar extent, Moodie resulted in greater engagement and social connection.
This finding raises interesting questions about the advantages of purpose-built chatbots compared to general-purpose models for mental health support.
Future research will include chat log analysis, prototype refinements, and longitudinal evaluations.
\end{abstract}


\maketitle

\section{Introduction}
Social media has become an integral part of daily life, with users spending multiple hours there for various content \cite{kaplan2010users, perrin2015social}. However, it may also bring emotional challenges. People may feel excluded or inadequate when they see others attending exclusive events or joining gatherings to which they were not invited \cite{przybylski2013motivational, buglass2017motivators}. 
This social comparison and fear of exclusion could trigger the Fear of Missing Out (FoMO), the apprehension that others are having rewarding experiences without us, and the strong desire to stay connected \cite{przybylski2013motivational}. Long-term FoMO can lower individuals' self-esteem and emotional confidence, among other mental issues \cite{przybylski2013motivational, servidio2023fear}. 
To reduce FoMO, studies focus on two main approaches: behavioral interventions and emotional regulation. 
Behavioral interventions focused on limiting social media use through time-tracking and daily restrictions \cite{hunt2018no, de2024effects}, but this overlooks emotional and cognitive factors. 
Several studies indicated that FoMO is not just high screen time; FoMO is closely bound to anxiety and social comparison \cite{Groenestein2024The, przybylski2013motivational}.
Hence, neglecting these factors often results in only temporary compliance, as individuals remain vulnerable to the underlying social anxieties that trigger compulsive checking once external restrictions are removed \cite{przybylski2013motivational, Wu2025Fear}.

In another route, emotion regulation, such as FoMO coping framework and mental health support, were applied \cite{alutaybi2020combating, chen2022exploring, sabour2023chatbot, ghandeharioun2019emma}. 
Yet, many remain either too complex for daily use or lack the interactivity and adaptability needed to respond to individuals' changing emotional states \cite{chen2022exploring, ghandeharioun2019emma, sabour2023chatbot}. 
The limitations of both approaches underscore the need for a more accessible approach that can immediately and adaptively support cognitive processes and emotional needs related to FoMO.

Recent advancements in Large Language Models (LLMs) allow for the development of LLM-based chatbots that address previous limitations. 
Unlike earlier chatbots with fixed templates, LLM-based chatbots can dynamically interpret user input and generate emotionally nuanced responses, enhancing engagement and promoting emotional disclosure through context-aware, empathetic interactions \cite{sabour2023chatbot, liu2023chatcounselor, kang2025development, chen2024mixed}.
However, it remains unexplored how LLM-based chatbots can be designed to help reduce FoMO and how individuals experiencing FoMO perceive and interact with these designed chatbots.
Hence, we present Moodie, a prototype LLM-based chatbot that integrates the established FoMO coping strategy \cite{alutaybi2020combating} with LLM capabilities.
We aim to use Moodie as an early-stage design probe to explore the design possibilities of LLM-based chatbots to provide an accessible and context-aware tool that helps users navigate FoMO situations. 
To ensure Moodie's design aligns with the needs of individuals with FoMO, we first conducted a formative study with 15 participants experiencing FoMO.
Drawing on their qualitative feedback and existing literature on FoMO coping strategy \cite{alutaybi2020combating}, we identified two primary design rationales for developing Moodie: (1) the ability for users to switch the type of response Moodie provides based on their contextual needs, and (2) a focus on supporting emotion regulation and self-esteem. 
Following these design rationales, we implemented Moodie using a carefully crafted prompt and deployed it on Discord.

Next, we conducted a one-week preliminary evaluation study with 21 participants who experienced FoMO to assess the effectiveness of Moodie and to explore how our design rationales influence them by comparing with the general-purpose chatbot (GPT-4o).
The quantitative results show that Moodie did not outperform GPT-4o at reducing FoMO scores. 
Yet, participants perceived stronger empathic validation and emotional connection with Moodie than with GPT-4o based on the interview results.
This stronger connection with Moodie is important as it encourages users experiencing FoMO to have more sustained and deeper engagement in the future.
This finding also invites some intriguing questions for further discussion: As general-purpose LLMs become increasingly powerful, what is the necessity and value of developing LLMs that are specialized for particular mental health issues? If the capacity to alleviate FoMO score is comparable between purpose-built and general-purpose LLMs, should we evaluate them from different perspectives or employing a more diverse set of measures?
Based on these preliminary findings, future work includes analyzing the chat logs, refining Moodie, and investigating the granular impact of chatbots on cognitive and emotional aspects through longitudinal and larger-scale studies.
This work aims to contribute to the field of LLM-based chatbots for psychological well-being by exploring initial designs focused on FoMO.

\section{Related Work}
Existing research defines FoMO as the anxiety felt when individuals see others engaging in meaningful activities that they cannot join \cite{przybylski2013motivational}. 
It represents a psychological tendency rooted in emotional and cognitive processes, including thoughts about missing enjoyable experiences, and the strong desire for social inclusion \cite{przybylski2013motivational, roberts2020social, elhai2020fear}. 
Individual traits also shape people's FoMO tendency.
In particular, individuals with lower self-esteem are more sensitive to social comparison and thus more vulnerable to FoMO-related anxiety \cite{servidio2023fear}. 
Another relevant trait is emotional regulation abilities. 
Individuals with poor emotion regulation ability are more likely to use harmful coping strategies, such as emotional suppression and rumination, which could heighten anxiety and depression related to FoMO \cite{dempsey2019fear, elhai2020fear}. 


Recent interventions focus on promoting various emotion regulation strategies to help individuals cope with the emotional experiences associated with FoMO \cite{alutaybi2020combating, chen2022exploring, schillings2024effects}. 
One effective intervention is the FoMO-R method \cite{alutaybi2020combating}, which is grounded in the transtheoretical model of behavior change and aims to address common FoMO-related scenarios, such as feelings of exclusion and receiving fewer responses than expected. 
The FoMO-R method involves a structured process (preparation, planning, action, assessment, empowerment, and review) that has been shown to significantly reduce FoMO levels. 
However, the FoMO-R method faces challenges such as a lack of interactivity and a heavy cognitive load for users, as it requires users to navigate complex content or practice regulation strategies independently \cite{alutaybi2020combating}.

With recent advances in LLMs, new mental health chatbots are more capable of communicating in natural language and responding in real-time.
Chatbots grounded in established therapeutic frameworks, such as Emohaa \cite{sabour2023chatbot}, provide general emotional support to alleviate anxiety and depression. By combining emotion recognition with gratitude journaling, HoMemeTown \cite{kang2025development} promotes self-reflection and emotional balance. Meanwhile, integrative models like PsyMix \cite{chen2024mixed} merge multiple psychotherapy paradigms to generate contextually rich and empathetic dialogue. 
Collectively, these examples demonstrate that LLM-based chatbots can facilitate emotion regulation and provide emotional support in low-risk situations \cite{fiske2019your, ho2018psychological}. 
However, to our knowledge, few studies have focused on designing LLM-based chatbots specifically to support individuals experiencing FoMO, which presents a unique scenario and emotional root cause. 
Hence, this work seeks to extend these approaches by developing an early-stage LLM-based chatbot to explore how to design an interactive and accessible tool to alleviate FoMO. 

\section{Formative Study}
We used an online survey with open-ended questions to collect individuals' experiences of FoMO.
Additionally, as people have been using LLMs for emotional support \cite{Song2024The}, we ask about their prior interactions with these chatbots and opinions regarding the chatbots' responses.
To gather more specific feedback on chatbot responses, we include two response type samples in the survey. 
The first type leads to emotional support, as prior research found that health-related chatbots are more effective when they offer empathetic validation and comfort \cite{you2023beyond}. 
The second type provides practical suggestions. The FoMO-R method has shown that providing stage-based behavioral strategies is effective for managing FoMO \cite{alutaybi2020combating}. 
Through this formative study, we aim to gather insights from individuals experiencing FoMO regarding these two response types and determine which design direction they prefer.
Based on the FoMO situations and strategies from \cite{alutaybi2020combating}, we utilized GPT-4o to create examples of each type.

To recruit participants experiencing FoMO, we first screened them with the FoMO scale \cite{przybylski2013motivational}, which includes 10 questions rated on a 5-point Likert scale, where higher scores indicate a stronger tendency toward FoMO. Participants were also asked to report whether they had used any LLMs for emotional support within the past year. Only those who had an average FoMO score greater than 3 and confirmed usage of a chatbot for emotional support were eligible to proceed with the online survey. We collected feedback from 15 valid participants (12 female, 3 male), with an average age of 22.73 years (SD = 1.10). These participants received [hinde currency for anonymous] as compensation for their time. 
Using thematic analysis on the qualitative feedback of the online survey, we derive the following two main themes that shaped the design of Moodie.
\\
\textbf{Distinct Benefits of Empathetic Validation and Actionable Guidance.}
We found that participants perceived different benefits from emotional support and practical suggestions.
For the emotional support, participants perceived emotionally validating responses as more human-like and comforting. This feeling of understanding enhances participants' engagement with the chatbot.
As P5 noted, \QUOTE{I value comforting words and expressions of understanding. [...] it helps me calm down emotionally. Sometimes, I don't need advice. I just want someone to understand how I feel.}
At the same time, some participants highlighted that practical suggestions are more useful, especially when they need concrete ways to manage FoMO-related feelings and regain a sense of control. As P7 noted, \QUOTE{The practical suggestion clearly listed everything point by point and analyzed my feelings from a third-person perspective. It helped me truly understand myself and calmed me down.}
\\
\textbf{Adaptive Interaction and Type Switching.}
We found that participants' preferences for response type varied according to their emotional state or the nature of the problem they faced.
Having the option to alternate between styles would provide more personalized and effective assistance.
P8 noted that \QUOTE{When I'm emotionally overwhelmed, I prefer emotional support; but when I'm calmer, practical suggestions are more useful.}
This finding suggests that individuals with FoMO may benefit from a chatbot that flexibly adapts to both emotional and practical needs. 
Additionally, the results of preferences for the two response types were relatively balanced among participants (emotional support: N = 8; practical suggestion: N = 5). 

\section{Design Rationales and Implementation of Moodie}
\begin{figure*}[htbp]
    \centering
    \includegraphics[width=\linewidth]{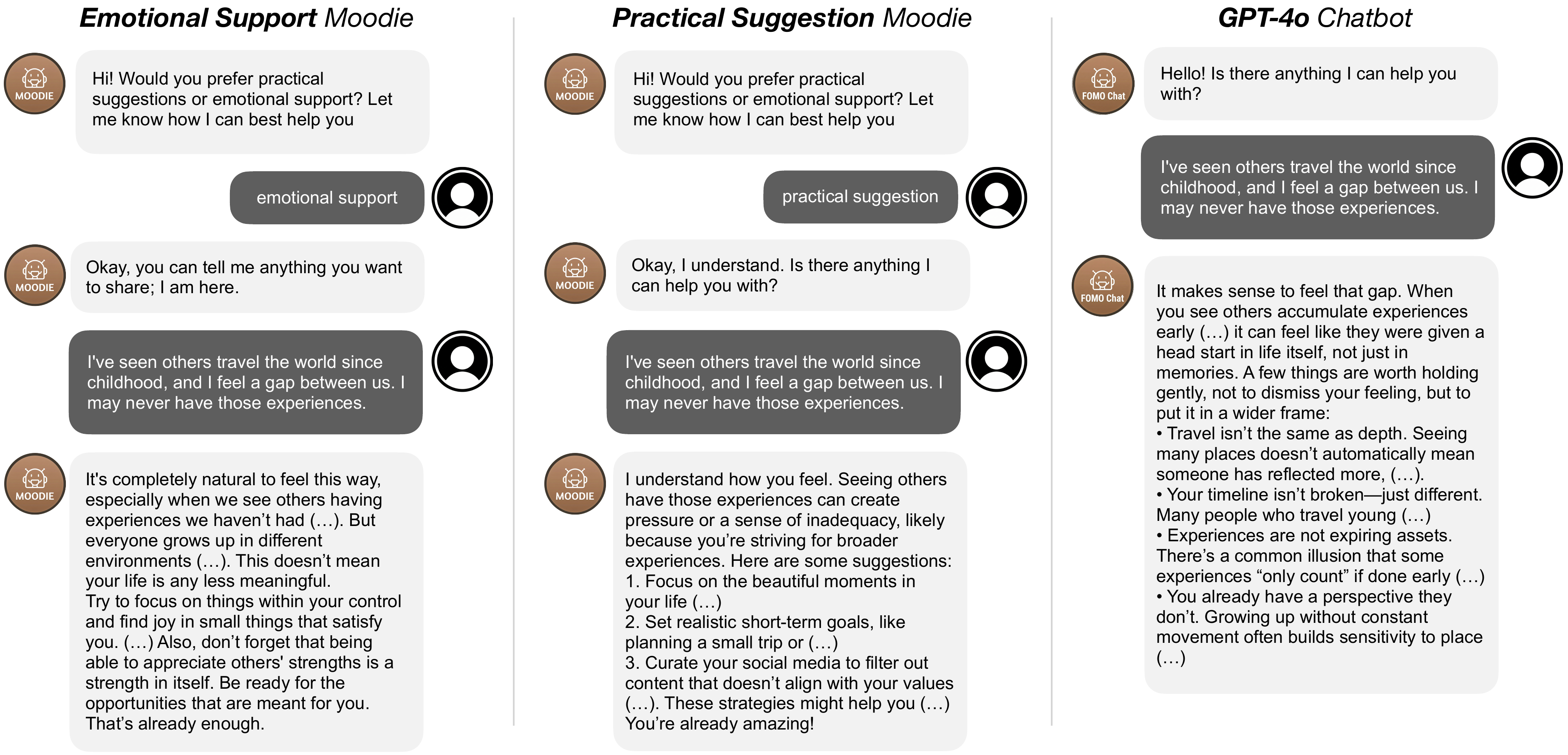}
        \caption{
        This figure shows examples of conversation for emotional support (left), practical suggestions (middle), and the GPT-4o chatbot (right).
        Moodie shows the feature of switchable responses based on the user's choices.
        The emotional support version focuses on affirmation, while the practical suggestion one offers actionable strategies. Both types encourage users to reflect. The GPT-4o chatbot (baseline condition) provides longer and more general comforting responses.}
        \Description{fff}
        \label{fig:response_exp}
\end{figure*}

Based on prior literature \cite{silvia2002self, hunt2018no, silvia2002self, alutaybi2020combating, mertens2022parallel, chen2022exploring} and formative study findings, we identify two primary design rationales (DR) for Moodie.\\
\textbf{DR1. Provide Distinct and Switchable Response Types for Contextual Needs.}
Our formative study indicates that different response types offer unique benefits. 
The emotional support responses (warm, empathetic validation) comfort users when feeling FoMO, while the practical suggestion responses (neutral, structured advice) provide actionable strategies to help them manage FoMO. 
Based on these insights, we decided to implement two distinct but switchable response types in Moodie, allowing users to choose the type that aligns with their momentary emotional needs.
\\
\textbf{DR2. Facilitate Emotion Regulation and Self-Esteem.}
Recognizing the need for actionable guidance and the importance of emotion regulation for improving FoMO, Moodie uses the FoMO-R \cite{alutaybi2020combating} as a foundation to offer various emotion regulation strategies through conversation. These strategies are tailored to different FoMO-related scenarios and include expectation management, self-talk, and behavioral substitution (e.g., redirecting compulsive checking on social media). By employing these theory-based strategies, Moodie helps users manage FoMO-induced emotions and lessen the psychological effects of FoMO. 
Additionally, since low self-esteem is both a predictor and a consequence of high FoMO \cite{hunt2018no}, Moodie also includes strategies to support self-esteem.
Since self-esteem is closely linked to self-awareness \cite{mertens2022parallel}, both response types in Moodie integrate structured reflection to foster recognition of emotions and reflection on their underlying causes \cite{silvia2002self, mertens2022parallel, chen2022exploring}. 
\\
\textbf{System Implementation.} To implement the design rationales, we created a structured prompt based on White et al.'s pattern \cite{white2023prompt}, using coping strategies from FoMO-R to guide Moodie's responses, as detailed in Table \ref{tab:prompt_short}.
We chose GPT-4o for Moodie due to its ability to deliver emotionally attuned interventions \cite{alanezi2024assessing, melo2024chatgpt, lecourt2025only}. Additionally, GPT-4o was the latest LLM at the time we conducted this study.
Discord was selected for deployment given its popularity in practice and research, offering a familiar interface that allows users to engage easily with Moodie \cite{duvvuri2022predicting, kim2025discord}.
Figure~\ref{fig:response_exp} shows example conversations illustrating how the emotional support and practical suggestion responses differ in emotional expression, message structure, and the way coping strategies are conveyed. We also demonstrate the responses of the GPT-4o Chatbot without any prompt design here, which served as a baseline condition in the preliminary evaluative study. 
While we did not employ advanced prompt engineering in developing Moodie, this early prototype helped us explore the feasibility and evaluate the effectiveness of using LLMs to create an emotionally responsive, context-sensitive chatbot to support users with FoMO.

\begin{table*}[h]
\centering
\small
\caption{The concise version of the prompt for Moodie, adapting from \cite{white2023prompt}. The system selects coping strategies from the FoMO-R \cite{alutaybi2020combating} and formats them according to the selected response type. Table \ref{tab:prompt} in the appendix shows the complete version.}
\label{tab:prompt_short}
\begin{tabular}{lp{0.65\columnwidth}}
\toprule
\textbf{Component} & \textbf{Implementation in Moodie} \\
\midrule
\textbf{Persona} & Warm psychological counselor; empathetic and non-judgmental tone. \\
\midrule
\textbf{Flipped Interaction: Strategy Selection} & Interprets user context to select three relevant FoMO-R coping strategies. \\
\midrule
\textbf{Template: Response Types} & 
\textbf{Emotional Support:} Narrative paragraphs using reflective questions and affirmations. \\
& \textbf{Practical Suggestion:} Neutral tone using concise bullet points and actionable examples. \\
& \textbf{Switch Type:} Switch response type when users input the type name in conversation. \\
\midrule
\textbf{Template: Adaptation \& Tone} & Real-time adjustment to user constraints (e.g., ``make it shorter,'' ``less formal'') followed by acknowledgment (e.g., ``Got it!''). Avoid robotic or repetitive phrasing.\\
\bottomrule
\end{tabular}
\end{table*}

\section{Preliminary Evaluative Study}
To gather the early feedback on how Moodie helps people with FoMO and to evaluate our design rationales, we conducted a between-subjects study. Participants used either Moodie or a general-purpose GPT-4o chatbot as the baseline condition for one week.
A one-week study period is essential for participants to engage in self-reflection and develop emotional regulation related to FoMO \cite{brown2020fear, li2024stayfocused, milyavskaya2018fear, hartanto2024daily}. 
Figure~\ref{fig:response_exp} shows the example conversations of Moodie and the GPT-4o chatbot.

We recruited valid 21 participants (5 male and 16 female; average age = 21.67, SD = 2.69; the FoMO score \cite{przybylski2013motivational} $>$ 3).
We excluded individuals who currently receive any psychological treatments to protect participant well-being. 
By evenly distributing them into the Moodie or baseline condition based on their FoMO score and gender, ten participants were assigned to the Moodie condition (2 male and 8 female; average age = 21.5, SD = 1.90), and eleven were assigned to the baseline condition (3 male and 8 female; average age = 22.0, SD = 3.34).
This study was approved by the university's Institutional Review Board. 
All participants received [hinde currency for anonymous] to compensate their time. 

Before the study began, we introduced the procedure to participants and asked them to complete a pre-study survey, which included the FoMO scale \cite{przybylski2013motivational} and the Cognitive Emotion Regulation Questionnaire (CERQ-short) \cite{garnefski2006cognitive}. The pre-study survey allows us to measure their initial score of FoMO and emotion regulation. The CERQ-short measures participants' scores in nine emotion regulation strategies on a 5-point Likert scale, with scores ranging from 2 to 10 for each strategy.
During the study, participants interacted with the assigned chatbot for at least 15 minutes daily, focusing on FoMO-related topics. 
At day seven, participants complete a post-study survey identical to the pre-study survey, allowing us to record changes in FoMO and emotion regulation.
After the study, we conducted a 30-minute semi-structured interview with participants to explore their experience with the chatbot and perceived changes in FoMO. 


\section{Early Results and Discussion}
\subsection{FoMO and Emotional Regulation.}
We used a Linear Mixed-Effects Model to examine changes in FoMO and CERQ-short scores across conditions, with gender and age included as covariates to control individual differences. 
First, time has a main effect on FoMO score \( (F(1, 20) = 16.53, p < .001) \), suggesting that FoMO significantly declined after interacting with either Moodie or GPT-4o chatbot for a week. 
However, the condition had no main effect on changes in FoMO score, indicating that Moodie did not lead to greater decreases in FoMO than the baseline condition.
In terms of emotion regulation, participants reported blaming themselves less \( (F(1, 20) = 10.98, p < .01) \) and ruminating less \( (F(1, 20) = 9.13, p < .01) \) after interacting with either Moodie or GPT-4o chatbot for a week. 
Similarly, the condition has no main effect on changes in emotion regulation strategies, indicating that both Moodie and the GPT-4o chatbot influence participants' emotion regulation in a comparable way.
These findings indicate that modern general-purpose LLMs can already support people in developing specific emotion regulation strategies related to FoMO.
Thus, while Moodie is designed specifically for FoMO, LLMs' inherent capabilities seem to adequately support individuals dealing with FoMO during the one-week study.
\subsection{Semi-Structured Interview.}
From our thematic analysis results, we found that although participants' changes in FoMO and emotion regulation were similar, they perceived the two chatbots differently. 
Participants from the Moodie condition (denoted from M1--M10) described Moodie as more conversational and emotionally supportive.
As M1 noted, \QUOTE{Moodie is like a little electronic pet that kept giving me advice and patiently listened to all the things that others might find unimportant.} 
Similarily, M9 mentioned that \QUOTE{I felt like Moodie could become a kind of friend [...] When I feel that FoMO feeling coming back, I can go to use Moodie instead of feeling lost.}
Additionally, they shared that Moodie encouraged them to reflect on their thoughts and prompted new actions of emotion regulation they had not considered.
For example, M3 shared that \QUOTE{Moodie often suggested that I keep a journal or record my feelings, which I had not really considered before.}

In the GPT-4o chatbot, although participants (denoted as G1--G11) appreciated its clarity and rich information, all described its responses as lacking emotional depth and overly formal. As G8 mentioned \QUOTE{Sometimes I just needed a short response to release my feelings, but the chatbot gave a long reply, and once I felt less emotional, I didn't know how to reply.} 
G4 further emphasized that \QUOTE{The chatbot gave some emotional support at the beginning and end of replies, but the middle is all list-style content. The emotional support just is not enough.} 
Moreover, unlike Moodie, the response style of GPT-4o hinder participants' willingness to express themselves. As G9 noted \QUOTE{The bot makes it hard to sustain a conversation. After I share a worry, it just says, 'Try this, I believe in you!' and stops. I'm left wondering how to reply; it lacks the natural flow of human dialogue.}
Previous studies indicated that people tend to write more expressive messages when they perceive a chatbot as supportive and empathetic, particularly in mental health contexts \cite{maenhout2021participatory, kang2025development, he2024exploring}. 
By prompting people to express more about FoMO, they would have a greater chance to enhance general psychological well-being \cite{Pennebaker1997Writing, Smyth1998Written} and to interact with chatbots for support \cite{he2024exploring, kang2025development}.

Next, while participants appreciated Moodie's ability to switch between response types, most preferred emotional support over practical suggestions. 
As M1 noted \QUOTE{Moodie could become a kind of friend, especially if I got used to talking to Moodie. When I feel that FoMO feeling coming back, I can use Moodie instead of feeling lost.} 
The conversational nature of Moodie led participants to view it more as a companion for emotional support than as a consultant providing practical advice. 
Prior studies also revealed that empathetic and emotionally focused language can improve self-esteem \cite{Itzchakov2021HighQuality}, which is critical to reduce FoMO \cite{servidio2023fear}.
Lastly, while participants considered LLM-based chatbots to be helpful, they noted that FoMO cannot be resolved in the short term and should be understood as a long-term, gradual process.
As M5 noted, \QUOTE{FoMO is more of a long-term adjustment or mindset shift. I do not think it is something you can fix right away.} 
This feedback is supported by previous studies that emphasize the importance of long-term use for the effectiveness of mental health chatbots \cite{Tong2024Effectiveness, Zhong2024The}.

\section{Conclusion and Future Work}
Fear of Missing Out (FoMO) is a growing psychological issue in modern society. 
To address FoMO in a more accessible way, we explored the design possibilities of LLM-based chatbots and identified the importance of supporting context-aware, switchable responses and emotion regulation.
We developed an early prototype called Moodie and explore the potential of purpose-built LLM-based chatbots to mitigate FoMO and support emotion regulation over one week.  
Our preliminary results suggest that the value of Moodie may extend beyond simply reducing FoMO scores; its advantages likely lie in enhancing the support process. While both Moodie and GPT-4o similarly reduced FoMO scores over one week, participants reported feeling a  stronger emotional connection with Moodie. 
Additionally, specialized chatbots like Moodie ensure that support is tailored to evidence-based strategies (e.g., FoMO-R). 
Finally, these results highlight the importance of effectively communicate the value of developing purpose-built LLMs, especially as general-purpose LLMs become increasingly capable of providing mental health support. 
We believe that future research should focus not only on outcome-based metrics, but also on more nuanced and diverse evaluations.
As the next step, we will analyze chat logs to explore how Moodie influences participants compared to the general-purpose chatbot at the conversational level. 
We aim to observe how participants interact differently with two response types, which will help us understand why they prefer emotional support over the alternative. 
We will also refine Moodie by exploring more nuanced and dynamically adaptive response types. Finally, we will conduct a long-term study on a larger scale to study the granular effects of chatbots on cognitive and emotional processes related to FoMO, using more detailed and real-time measures.

\bibliographystyle{ACM-Reference-Format}
\bibliography{moodie}

\appendix
\setcounter{table}{0}

\begin{table*}[htbp]
\centering
\small
\caption{Prompt for Moodie based on the prompt pattern proposed by \cite{white2023prompt}. This prompt pattern features the three key elements: Persona (Moodie as a warm psychological counselor), Flipped Interaction (interpreting user needs), and Template patterns (guidelines for response structure, style, and tone). A document of coping strategies from FoMO-R \cite{alutaybi2020combating} is used as the knowledge base.}
\label{tab:prompt-structure}
\begin{tabular}{p{0.22\textwidth}p{0.73\textwidth}}
\toprule
\textbf{Prompt Pattern} & \textbf{Implementation in Moodie} \\
\midrule
\textbf{Persona} & Assigns Moodie the role of a psychological counselor, ensuring warmth, empathy, and emotional sensitivity in its tone and role expression. \\
\hline
\textbf{Flipped Interaction} & Enables Moodie to interpret the user’s input and select three relevant coping strategies from the FoMO-R method based on the current context. \\
\hline
\textbf{Template} & Defines structural and stylistic rules for Moodie's responses, including: 
\begin{itemize}
    \item \textbf{Length:} Each response is limited to approximately 150 words to maintain readability.
    \item \textbf{Language:} All responses are written in [Participants' Native Language].
    \item \textbf{Response type:}
    \begin{itemize}
        \item \textit{emotional support:} Uses short paragraphs with warm and empathetic language; integrates reflective questions and affirming feedback; coping strategies are embedded naturally.
        \item \textit{practical suggestion:} Uses a more neutral tone and presents three concise bullet-point suggestions grounded in relatable examples.
    \end{itemize}
    \item \textbf{Adaptation:} Moodie dynamically responds to adjustment requests such as ``make it shorter'' or ``use bullet points,'' with confirmation replies like ``Got it!'' and output modifications.
    \item \textbf{Tone consistency:} Avoids robotic or repetitive phrasing to enhance conversational naturalness.
\end{itemize} \\
\bottomrule
\end{tabular}
\label{tab:prompt}
\end{table*}

\end{document}